\documentclass[twocolumn,aps]{revtex4-1}

\pdfpageattr {/Group << /S /Transparency /I true /CS /DeviceRGB>>}

\usepackage{graphicx} 
\usepackage{tikz}

\DeclareSymbolFont{lettersA}{U}{txmia}{m}{it}
\DeclareMathSymbol{\real}{\mathord}{lettersA}{"92}
\DeclareMathSymbol{\field}{\mathord}{lettersA}{"83}

\begin{document}

\title{Quantifying the effects of local many-qubit errors and non-local two-qubit errors on the surface code}

\author{Austin G. Fowler$^{1,2}$, John M. Martinis$^1$}

\affiliation{$^1$Department of Physics, University of California, Santa Barbara, California 93106,
USA \\
$^2$Centre for Quantum Computation and Communication
Technology, School of Physics, The University of Melbourne, Victoria
3010, Australia}

\date{\today}

\begin{abstract}
Topological quantum error correction codes are known to be able to tolerate arbitrary local errors given sufficient qubits. This includes correlated errors involving many local qubits. In this work, we quantify this level of tolerance, numerically studying the effects of many-qubit errors on the performance of the surface code. We find that if increasingly large area errors are at least moderately exponentially suppressed, arbitrarily reliable quantum computation can still be achieved with practical overhead. We furthermore quantify the effect of non-local two-qubit correlated errors, which would be expected in arrays of qubits coupled by a polynomially decaying interaction, and when using many-qubit coupling devices. We surprisingly find that the surface code is very robust to this class of errors, despite a provable lack of a threshold error rate when such errors are present.
\end{abstract}

\maketitle

Many different approaches to achieving reliable quantum computation are under investigation \cite{Naya08,Bone12,Gott13,Bomb13,Brav13b}. The current most practical known approach, the Kitaev surface code \cite{Brav98,Denn99}, calls for a 2-D array of qubits with nearest neighbor interactions and a universal set of quantum gates with error rates below an approximate threshold of 1\% \cite{Raus07,Raus07d,Fowl12f}. Superconducting qubits with error rates at the surface code threshold now exist \cite{Bare13}.

There is extensive prior work showing the existence of a threshold error rate when arbitrary quantum error correction codes are subjected to a wide variety of noise models, including algebraically decaying two-body correlated noise \cite{Ahar06}, Gaussian non-Markovian noise \cite{Ng09}, and arbitrarily many-body correlated noise \cite{Pres13}. In this work we focus on the simulated performance of the surface code below threshold.

To date, when the surface code has been simulated, quantum gates have only had the potential to introduce errors on the qubits they manipulated directly. In reality, manipulating any given qubit may disturb the state of a large number of surrounding qubits. Not all types of disturbance are particularly dangerous. Small random or systematic rotations of surrounding qubits lead only to independent random errors. Only \emph{correlated} many-qubit errors deserve specific attention. This distinction is discussed in detail in Section~\ref{corr}. In this work we present a detailed study of precisely how well the surface code can handle this class of correlated errors.

Another important class of errors that has not received attention to date is those that would arise in an array of qubits interacting directly with one another via a polynomially decaying interaction such as the Coulomb or magnetic dipole interaction, or via a device coupling to many qubits. Pairs of qubits initially antiparallel can both flip without changing the energy of the total system. Two-qubit errors can therefore appear on widely separated qubits. We also present a detailed study of this class of correlated errors.

The discussion is organized as follows. In Section~\ref{corr}, the meaning of independent and correlated errors is discussed in detail. In Section~\ref{sc}, the surface code is briefly reviewed, our method of modeling local many-qubit errors is described, and simulation results of this case are presented. In Section~\ref{2q}, our method of modeling non-local two-qubit errors is described, and simulation results presented. Section~\ref{conc} concludes.

\section{Independent and correlated errors}
\label{corr}

\begin{figure}
\begin{center}
\includegraphics[width=75mm]{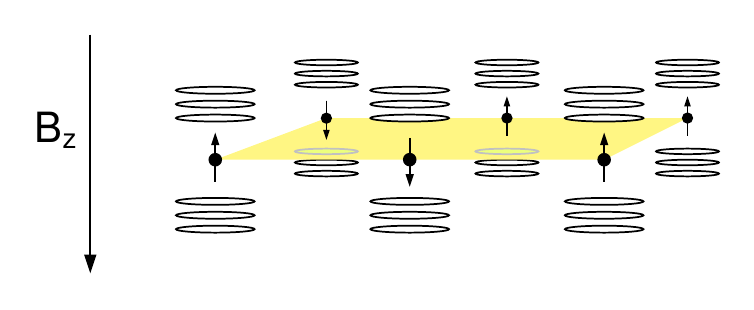}
\end{center}
\caption{(Color online) Hypothetical quantum computer architecture consisting of mobile spins on a cold substrate, each spin with its own hypothetical solenoid for readout and single-qubit gates, with two-qubit gates achieved by bringing neighboring spins closer together to increase the strength of the magnetic dipole interaction. This example has features in common with many physical architectures under investigation, especially those based on superconducting qubits.}\label{arch}
\end{figure}

Before presenting a study of correlated errors, it is worth discussing exactly what is and what is not a correlated error. For illustrative purposes, we center the discussion around a hypothetical quantum computer consisting of a 2-D array of mobile spins on a cooled substrate. A global magnetic field $B_z$ sets the energy difference between $|0\rangle$ and $|1\rangle$. Local solenoids above and below the default location of each spin provide localized AC and DC fields to drive arbitrary single-qubit rotations. Pairs of spins are moved into close proximity to raise the strength of the magnetic dipole interaction and implement two-qubit entangling gates. For simplicity, we also imagine the solenoids can be used, when desired, as sensitive magnetic field detectors for qubit readout. See Fig.~\ref{arch}. This example maps well to architectures based on superconducting qubits \cite{Bare13a}, spin qubits \cite{Holl06}, and quantum dots \cite{Loss98}, and has features common to architectures based on ion traps \cite{Kiel02}, optical lattices \cite{Bren99}, and many others.

We now consider various error sources, and whether they are, or are not, correlated error sources. Firstly, we consider small fluctuations in the global magnetic field $B_z$, which will lead to small undesired systematic $Z$ rotations on all qubits. At first glance, this may seem like the ultimate correlated error. However, provided the fluctuations are small and error detection is frequent, each individual small angle $Z$ rotation will just look like a small probability of a $Z$ error on each qubit. When performing error detection, most of the time no errors will be detected, as unwanted phase rotations will be removed by observation the majority of the time. This is a special case of the quantum Zeno effect \cite{Misr77}. Any detected errors will appear random and independent. A global fluctuating field leads to a correlated \emph{probability} of error $p$ on all qubits, but the errors themselves will not be correlated, and the probability of errors from this noise source on any given pair of qubits will be $p^2$. Note that it is critical that the fluctuations are small and error detection frequent --- for example if a global $\pi$ rotation accumulates this will indeed lead to a global correlated error.

Secondly, consider crosstalk when driving a single-qubit gate. Under the assumption of widely separated spins and small solenoids, each solenoid will look like a magnetic dipole and the field seen by other spins will decay cubically with separation. The driving field will therefore induce cubically decaying small-angle rotations in all spins in the computer. For the same reason that small fluctuations in the global field $B_z$ do not lead to correlated errors, this polynomially decaying crosstalk will also not lead to correlated errors. Provided the total error seen by any given qubit as a result of the sum of all crosstalk from all other actively manipulated qubits remains small, errors seen by the quantum error detection machinery will remain independent and sufficiently rare to be correctable.

Thirdly, consider the possibility that our hypothetical quantum computer is unshielded and located near an infrequent but energetic radiation source. Consider a hypothetical energetic particle that locally strongly heats the substrate on impact, but otherwise causes no physical degradation of the system. Imagine that the heating thermally randomizes spins in some neighborhood of the impact, with the neighborhood size proportional to the energy of the impact, and the probability distribution of increasingly energetic impacts decaying exponentially. Suppose furthermore that the cooling power per unit area of the substrate is sufficiently high to remove the excess heat in a small constant amount of time. This hypothetical scenario would lead to spatially correlated large-area errors with larger areas exponentially suppressed. Noise of this generic form shall be considered in Section~\ref{sc}. This section also considers the possibility of polynomial suppression of larger area errors.

Fourthly, consider direct magnetic dipole spin-spin interactions. A pair of antiparallel spins can spontaneously flip with no increase or decrease in energy of the system. The probability of this occurring is proportional to the interaction strength, which decays cubically. Pairwise noise of this form is two-body correlated noise. Note that each pairwise noise event requires the exchange of a virtual photon, so multiple pairwise noise events are random and uncorrelated. We shall consider noise of this generic form in Section~\ref{2q}.

Finally, imagine that spins are sufficiently separated to make the direct dipole-dipole interaction negligible, however there are elements in the physical construction that behave like inductive loops around each column of spins. These could be control lines or long-range qubit-qubit coupling elements. Now any pair of initially antiparallel spins in a given column can flip. Such a noise source would not be suppressed with increasing qubit separation. This form of noise shall also be considered in Section~\ref{2q}.

Undoubtedly other forms of noise could be considered, however we feel that the four correlated error classes listed above, namely 1) large-area exponentially decaying, 2) large-area polynomially decaying, 3) arbitrary qubit pairs polynomially decaying, 4) qubit pairs in columns non-decaying, cover the vast majority of basic behaviors likely to be found in physical devices. We would be happy to extend our work to cover other error classes of interest to the community, and welcome suggestions.

\section{Surface code performance with local many-qubit errors}
\label{sc}

\begin{figure}
\begin{center}
\begin{tikzpicture}
    \node[anchor=south west,inner sep=0] at (0,0) {\includegraphics[width=85mm, viewport=60 60 545 430, clip=true]{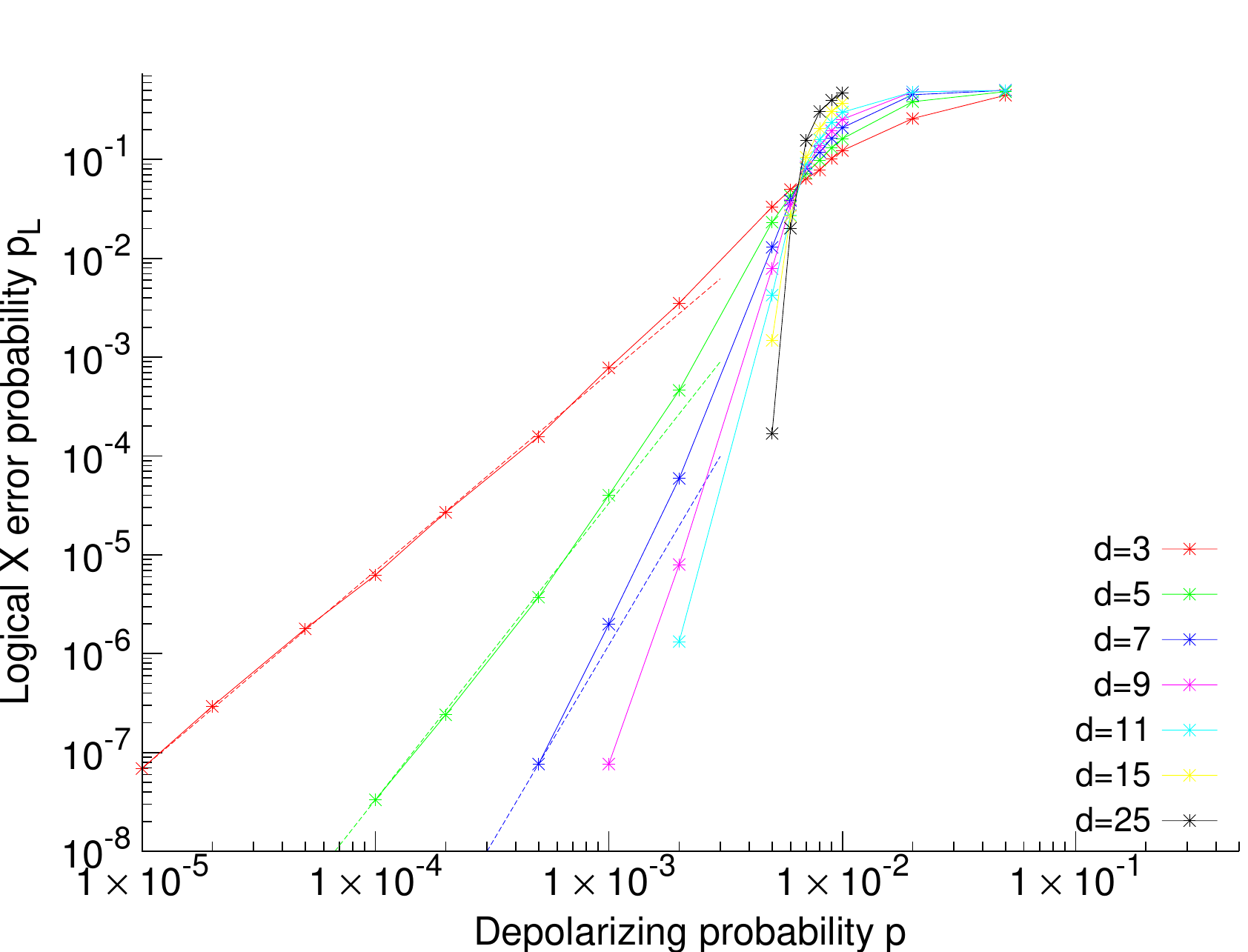}};
    \node [below right] at (1.5,6) {no correlated errors};
\end{tikzpicture}
\end{center}
\caption{(Color online) Probability of logical $X$ error per round of fault-tolerant error detection $p_L$ as a function of the depolarizing error probability $p$ for a range of distances $d=3, \ldots, 25$ when exploiting knowledge of correlations between $X$ and $Z$ errors. Referring to the left of the figure, the distance increases top to bottom. Quadratic, cubic, and quartic lines (dashed) have been drawn through the lowest distance 3, 5, 7 data points obtained, respectively.}\label{logx_ft_c}
\end{figure}

For our purposes, a distance $d$ surface code is simply a $(2d-1)\times (2d-1)$ 2-D array of qubits capable of protecting a single qubit of data by periodically executing a particular quantum circuit designed to detect errors \cite{Fowl12f}. If we assume that each quantum gate in the periodic circuit has an error rate $p$, then given a distance $d$ surface code we can use simulations to calculate the probability of a logical error per round of error detection $p_L$, namely the probability $p_L$ that we fail to protect the single qubit of data distributed across the lattice of qubits. Fig.~\ref{logx_ft_c} shows $p_L$ as a function of $p$ and $d$ using asymptotically optimal error suppression techniques \cite{Fowl13g}. This is our baseline performance. Introducing large-area errors will degrade this performance.

\begin{figure}
\begin{center}
\includegraphics[width=55mm]{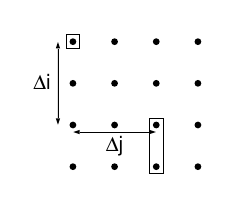}
\end{center}
\caption{Each dot represents a qubit. If a quantum gate is applied to the two qubits within the vertical rectangle, the qubit within the square is said to have $\Delta i=2$ and $\Delta j=2$, namely the minimum $(i, j)$ coordinate differences with any qubit acted on by the gate.}\label{didj}
\end{figure}

Consider Fig.~\ref{didj}, which defines two quantities $\Delta i$, $\Delta j$ that have meaning during the application of a quantum gate and will enable us to define our error models. We shall consider two particularly severe models of many-qubit errors, each with a single tunable parameter $n$ determining its strength. Unlike in Section~\ref{corr} where large-area errors were motivated by a particle impact example, we shall associate such errors with the every application of every quantum gate.

When applying a gate with error rate $p$, a single random number $x$ is generated. If $x<p$, the qubits involved in the gate will suffer random equally likely Pauli errors (with no chance of an identity error). Every other qubit in the surface code will suffer random equally likely errors $I$, $X$, $Y$, $Z$ if at the location of the qubit $x<p/n^{\Delta i+\Delta j}$ (exponential model) or $x<0.1p/r^n$, where $r=\sqrt{\Delta i^2+\Delta j^2}$ (polynomial model).

The motivation behind the exponential model's use of a non-Euclidean metric is qubits with a negligible direct qubit-qubit interaction that instead must be coupled via physical devices that are themselves non-interacting. In this scenario, qubits are physically well separated. The hypothetical energetic particle discussed in Section~\ref{corr} should be imagined as significantly raising the temperature or photon count of a specific component, and each successive device should provide additional isolation leading to Manhattan distance exponential suppression of the unwanted effects. The polynomial model is motivated by qubits that are closely spaced with thermal errors radiating through the substrate. All gates, including initialization, Hadamard, CNOT, measurement, and identity, are assumed to have a non-zero probability of suffering from such large-area errors during their implementation.

\begin{figure*}
\begin{center}
\begin{tikzpicture}
    \node [anchor=south west, inner sep=0] at (0,7) {\includegraphics[width=85mm, viewport=60 60 545 430, clip=true]{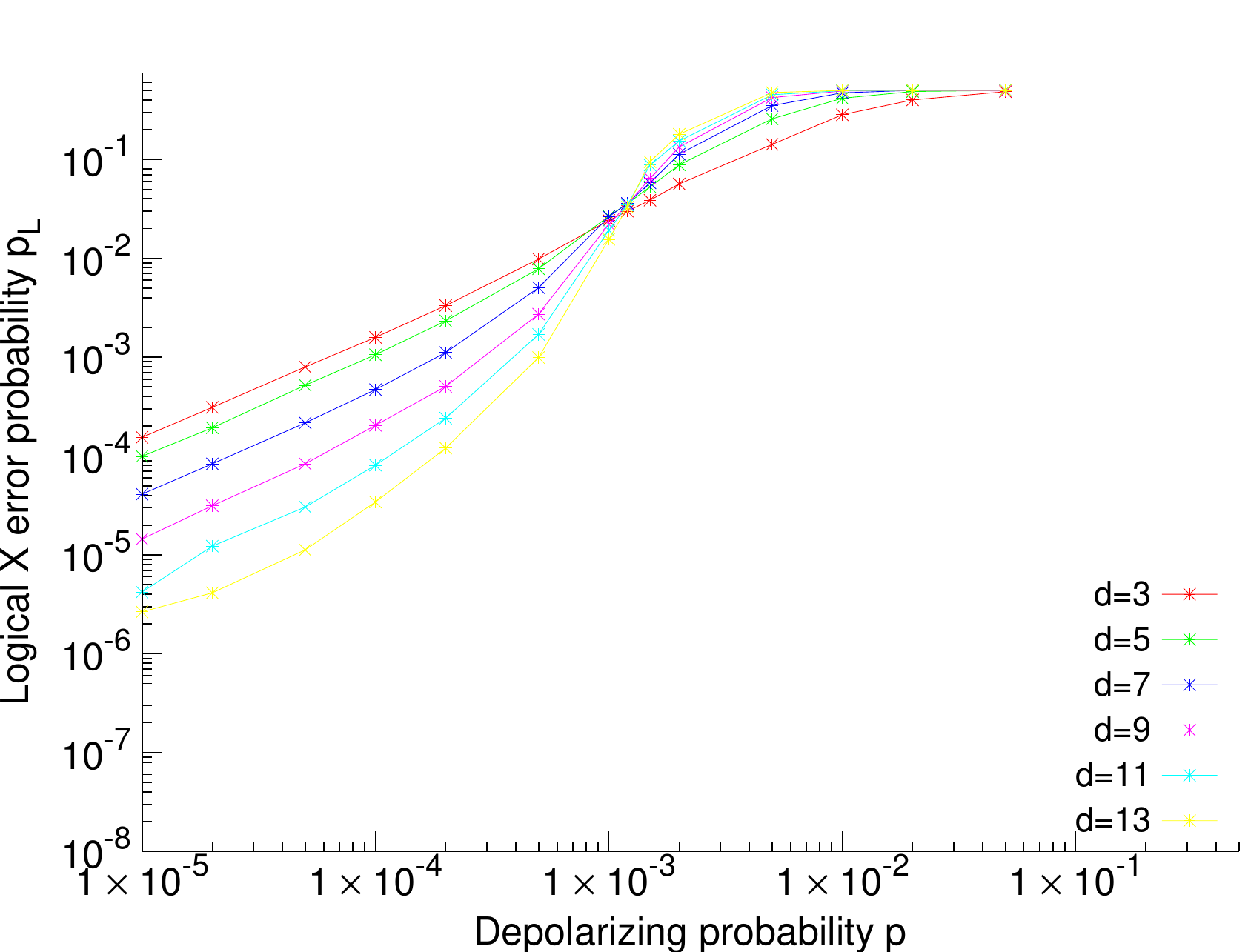}};
    \node [below right] at (0,13) {a)};
    \node [below right] at (1.5,13) {$n=2$};

    \node [anchor=south west, inner sep=0] at (9.5,7) {\includegraphics[width=85mm, viewport=60 60 545 430, clip=true]{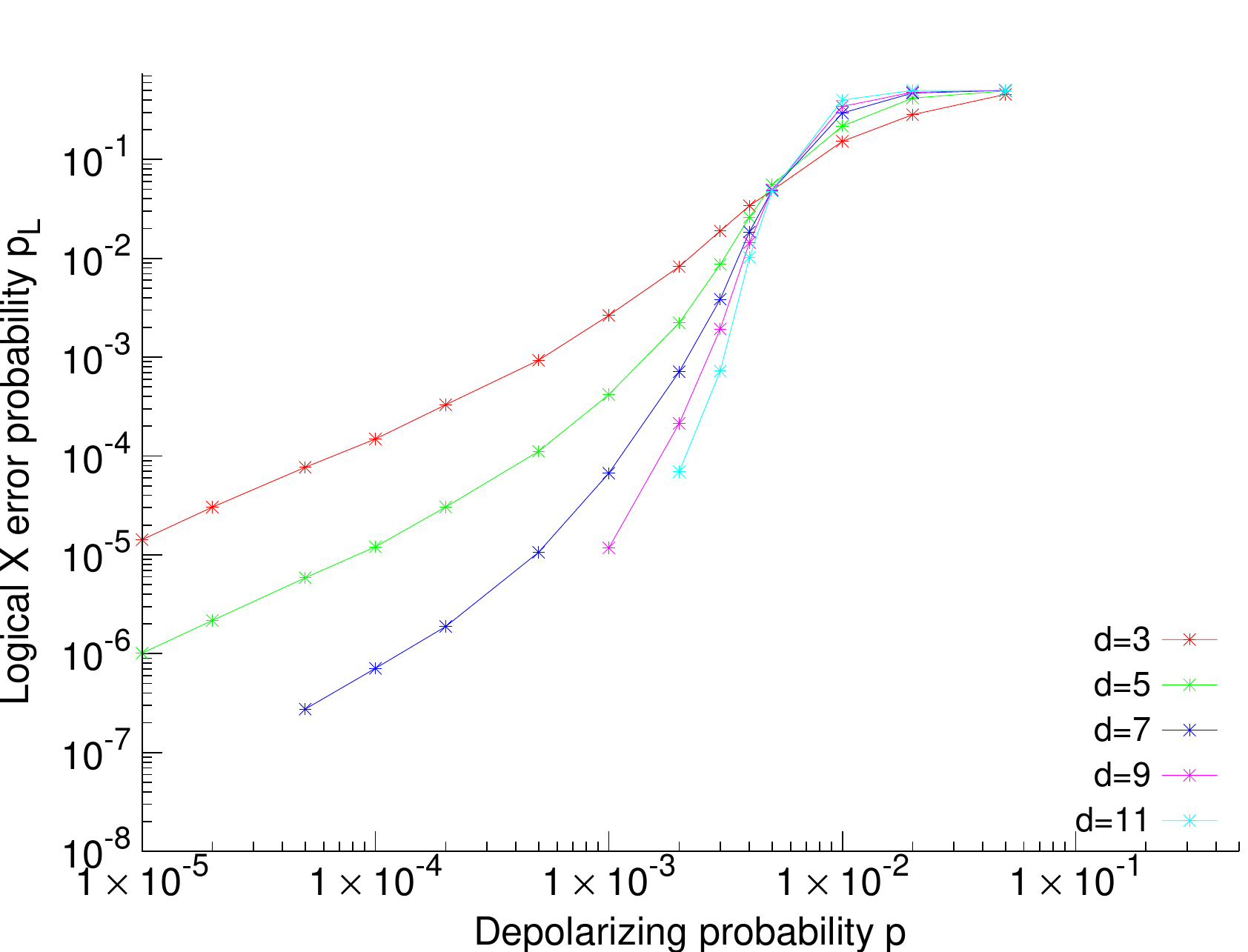}};
    \node [below right] at (9.5,13) {b)};
    \node [below right] at (11,13) {$n=10$};

    \node [anchor=south west, inner sep=0] at (0,0) {\includegraphics[width=85mm, viewport=60 60 545 430, clip=true]{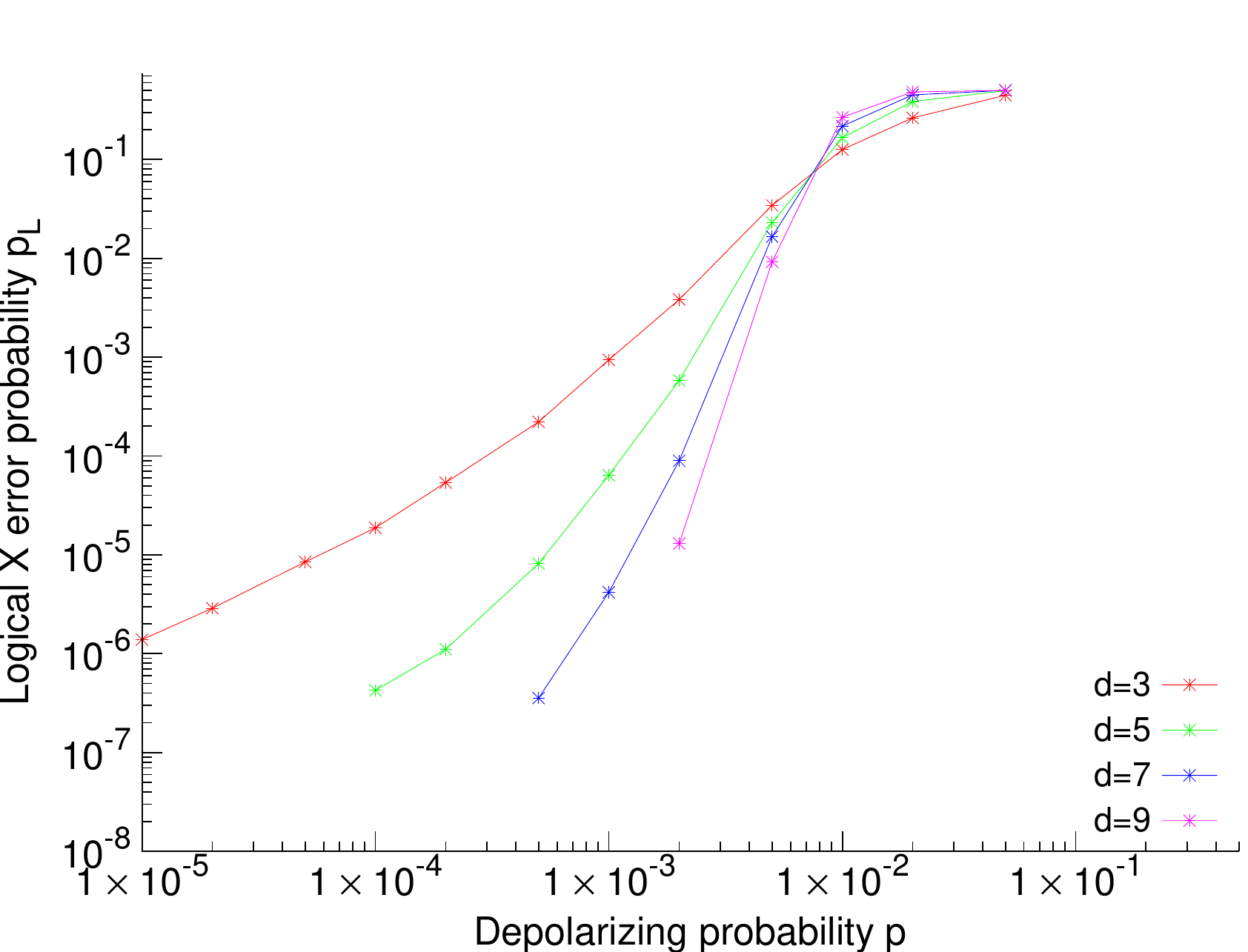}};
    \node [below right] at (0,6) {c)};
    \node [below right] at (1.5,6) {$n=100$};

    \node [anchor=south west, inner sep=0] at (9.5,0) {\includegraphics[width=85mm, viewport=60 60 545 430, clip=true]{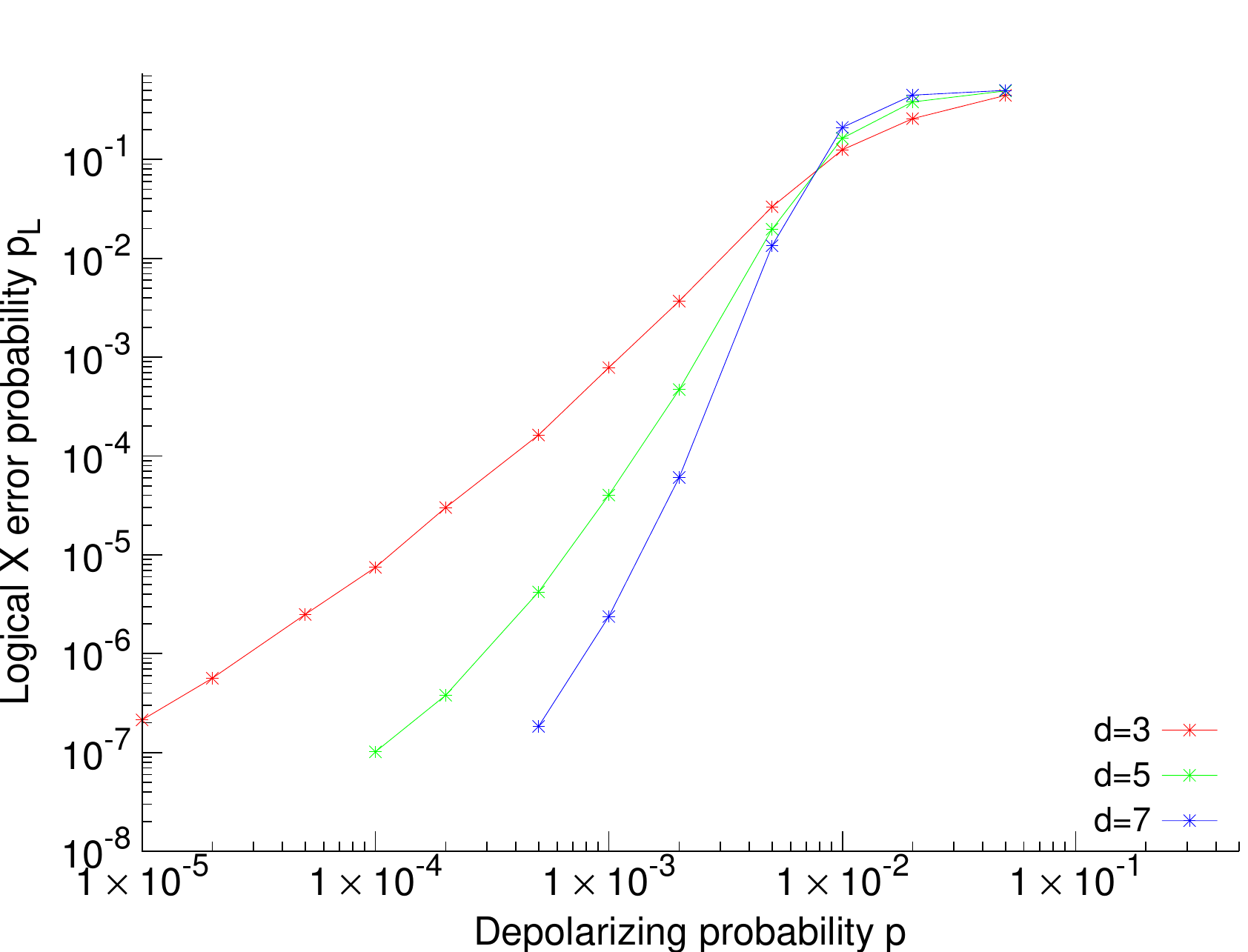}};
    \node [below right] at (9.5,6) {d)};
    \node [below right] at (11,6) {$n=1000$};
\end{tikzpicture}
\end{center}
\caption{(Color online) Probability of surface code logical $X$ error per round of error detection for various code distances $d$ and physical error rates $p$ when increasingly large area errors are exponentially suppressed by a factor of (top left) $n=2$, (top right) $n=10$, (bottom left) $n=100$, (bottom right) $n=1000$. Referring to the left of each graph, distance increases top to bottom. It can be seen (top right) that even when $n=10$, meaning a single-qubit gate with error rate $p$ has a probability $p/10$ of triggering a correlated plus-shaped 5-qubit error, and $p/100$ of triggering a correlated diamond-shaped 13-qubit error, and so on, and two-qubit gates similarly trigger higher weight errors, that robust and efficient exponential suppression of logical error can still be achieved.}\label{exp}
\end{figure*}
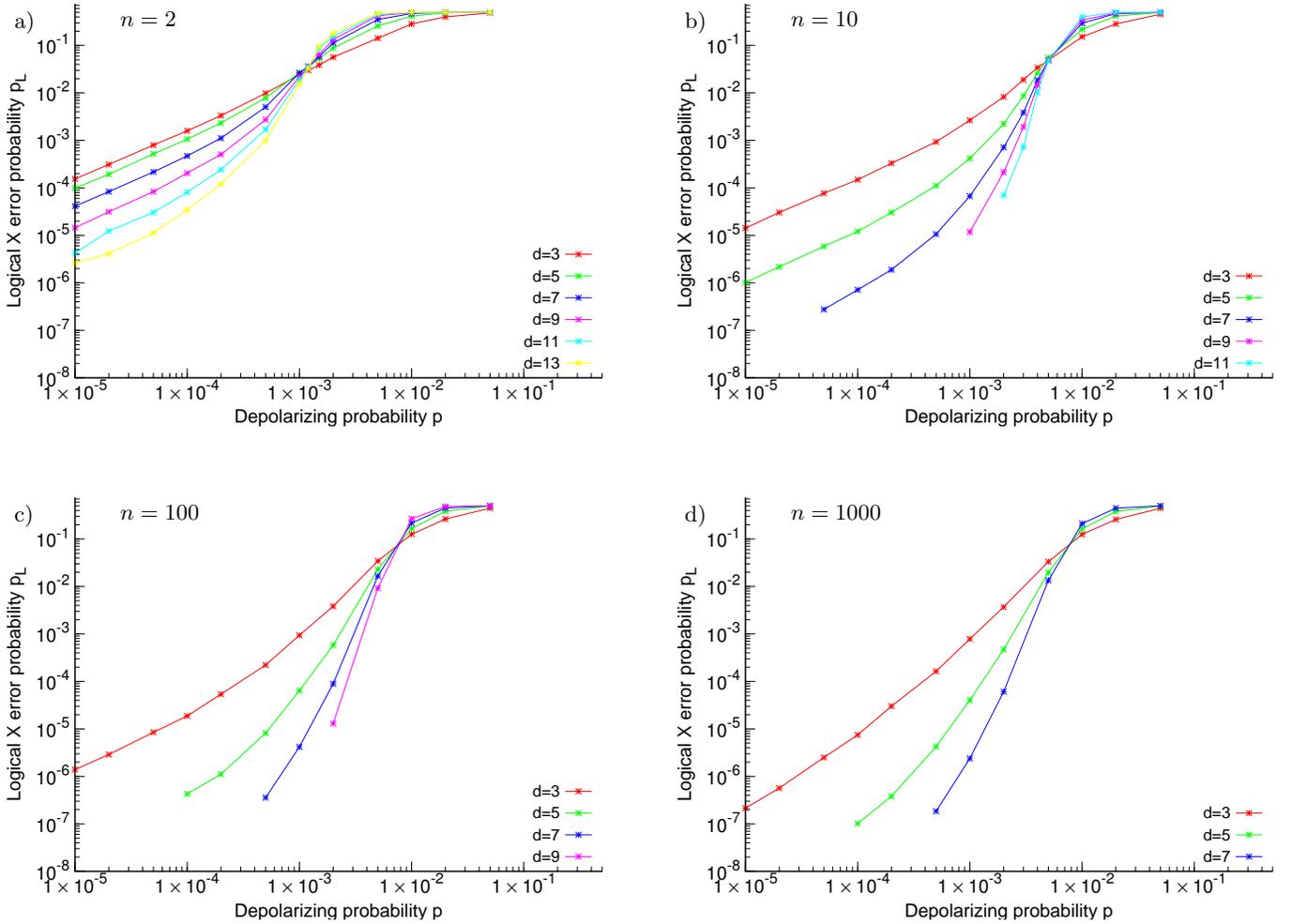

Figure~\ref{exp} shows the performance of the surface code with exponential model large-area errors and $n=2$, 10, 100, and 1000. It can be seen that performance is still measurably degraded even for $n=1000$, however strong exponential suppression of logical error at fixed $p$ can still be achieved even for $n=10$. To be quantitative, at an operating error rate of $p=10^{-3}$, in the absence of large-area errors (Fig.~\ref{logx_ft_c}), a distance $d=7$ surface code achieves a logical error rate per round of error detection of $p=2.0\times 10^{-6}$. For $n=1000$, this is degraded to $p=2.4\times 10^{-6}$. This level of degradation would have negligible practical impact, with very slightly larger code distances required to compensate. Even for $n=10$, where the logical error rate is degraded to $p=6.7\times 10^{-5}$, the degradation can be fully compensated by using a larger $d=11$ code, leading to an approximate factor of $(11/7)^2\sim 2.5$ additional qubits, independent of the size of the quantum computation protected in this manner. A factor of 2.5 overhead is significant but not excessively onerous, and we therefore claim that even quite moderate exponential suppression of large-area errors is tolerable in a practical manner when using the surface code.

A striking difference between Fig.~\ref{logx_ft_c} (no large-area errors) and Fig.~\ref{exp} (large-area errors) is the linear suppression of logical error in the latter for low values of $p$ at a fixed code distance $d$ as $p$ is reduced further. This is due to the fact that any single error has the potential to cause a logical error, and at low values of $p$ multiple temporally nearby gate errors become unlikely and the dominant logical error process becomes single large-area errors. Note that at fixed low $p$, logical error suppression is still exponential with increasing $d$.

\begin{figure*}
\begin{center}
\begin{tikzpicture}
    \node [anchor=south west, inner sep=0] at (0,7) {\includegraphics[width=85mm, viewport=60 60 545 430, clip=true]{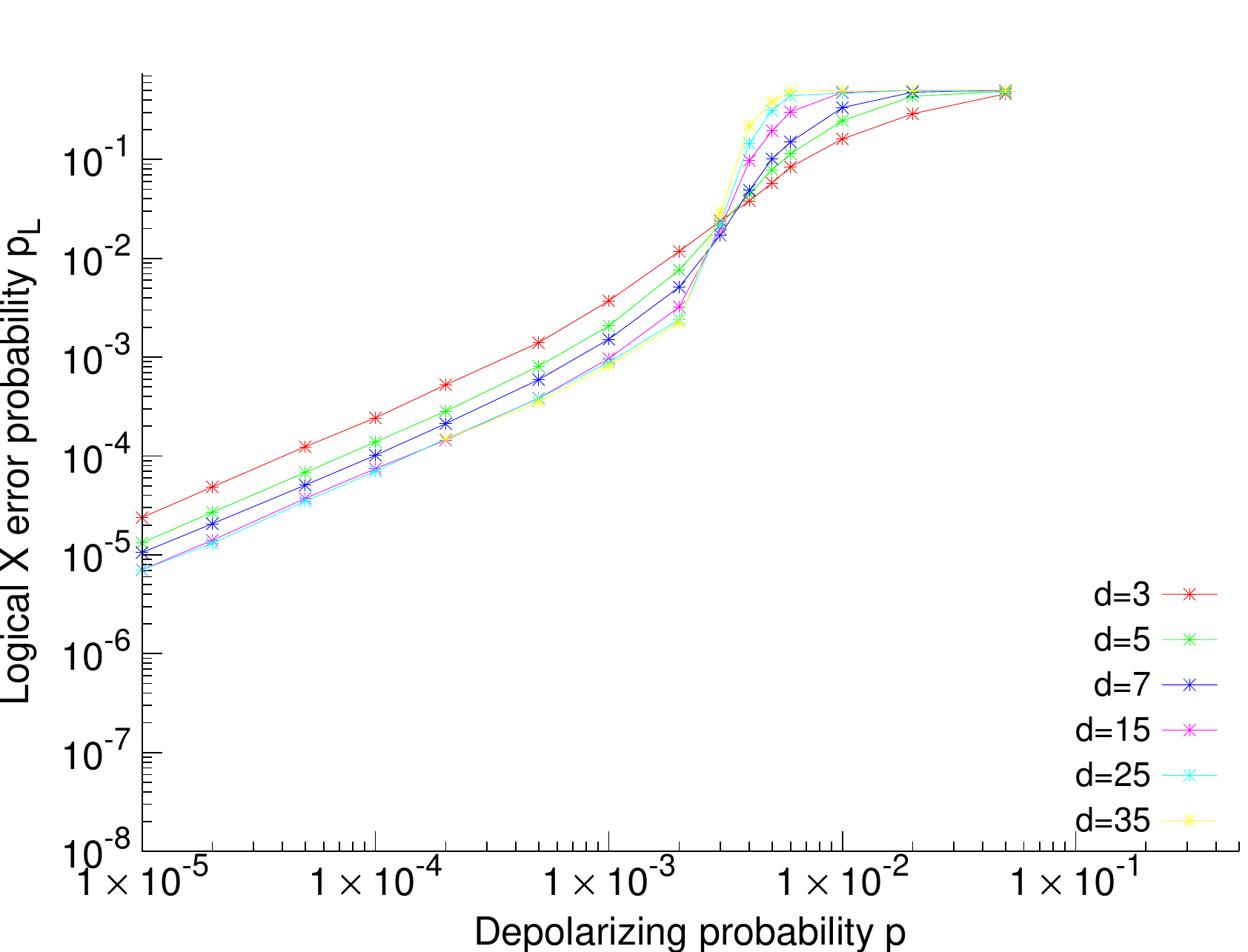}};
    \node [below right] at (0,13) {a)};
    \node [below right] at (1.5,13) {$n=2$};

    \node [anchor=south west, inner sep=0] at (9.5,7) {\includegraphics[width=85mm, viewport=60 60 545 430, clip=true]{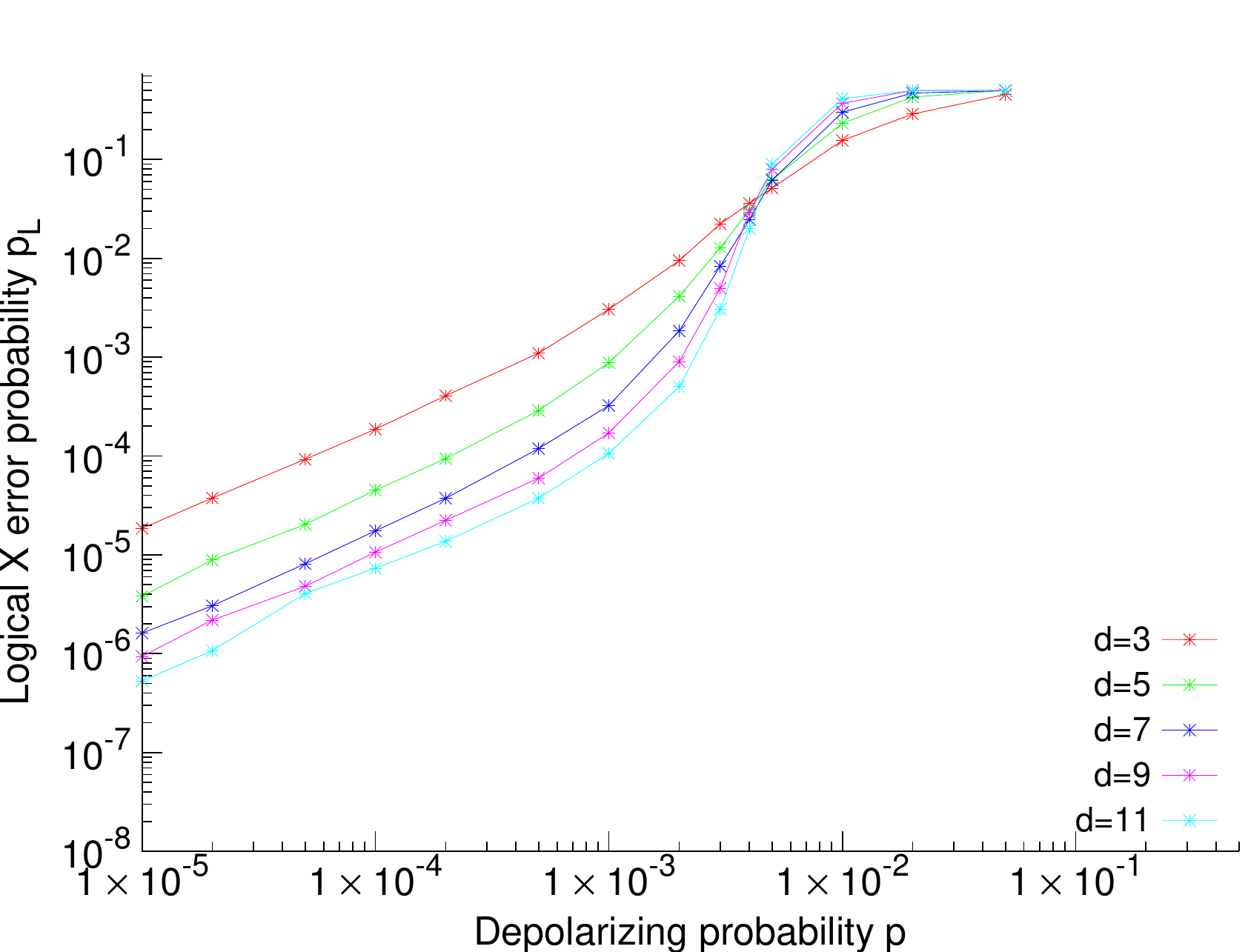}};
    \node [below right] at (9.5,13) {b)};
    \node [below right] at (11,13) {$n=3$};

    \node [anchor=south west, inner sep=0] at (0,0) {\includegraphics[width=85mm, viewport=60 60 545 430, clip=true]{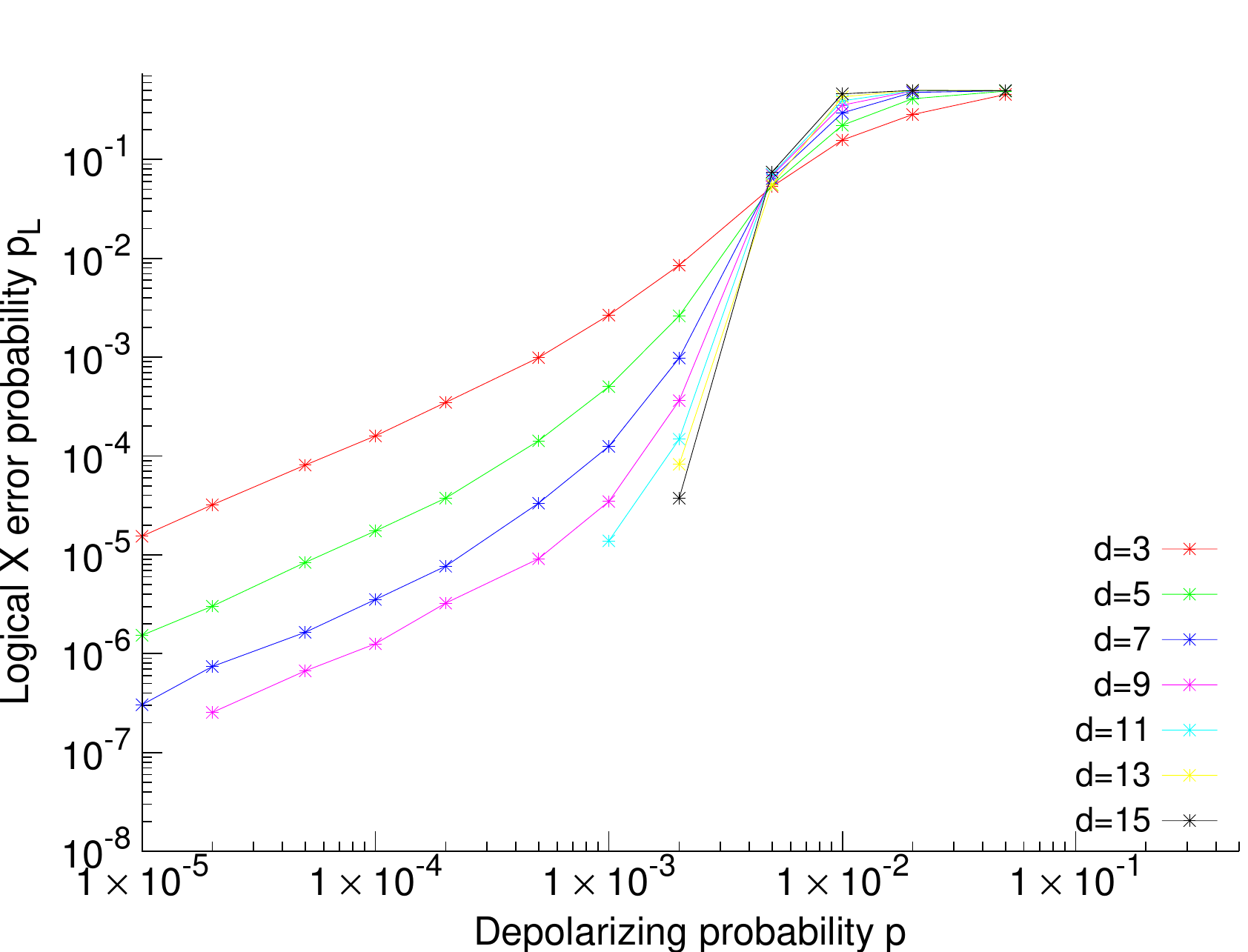}};
    \node [below right] at (0,6) {c)};
    \node [below right] at (1.5,6) {$n=4$};

    \node [anchor=south west, inner sep=0] at (9.5,0) {\includegraphics[width=85mm, viewport=60 60 545 430, clip=true]{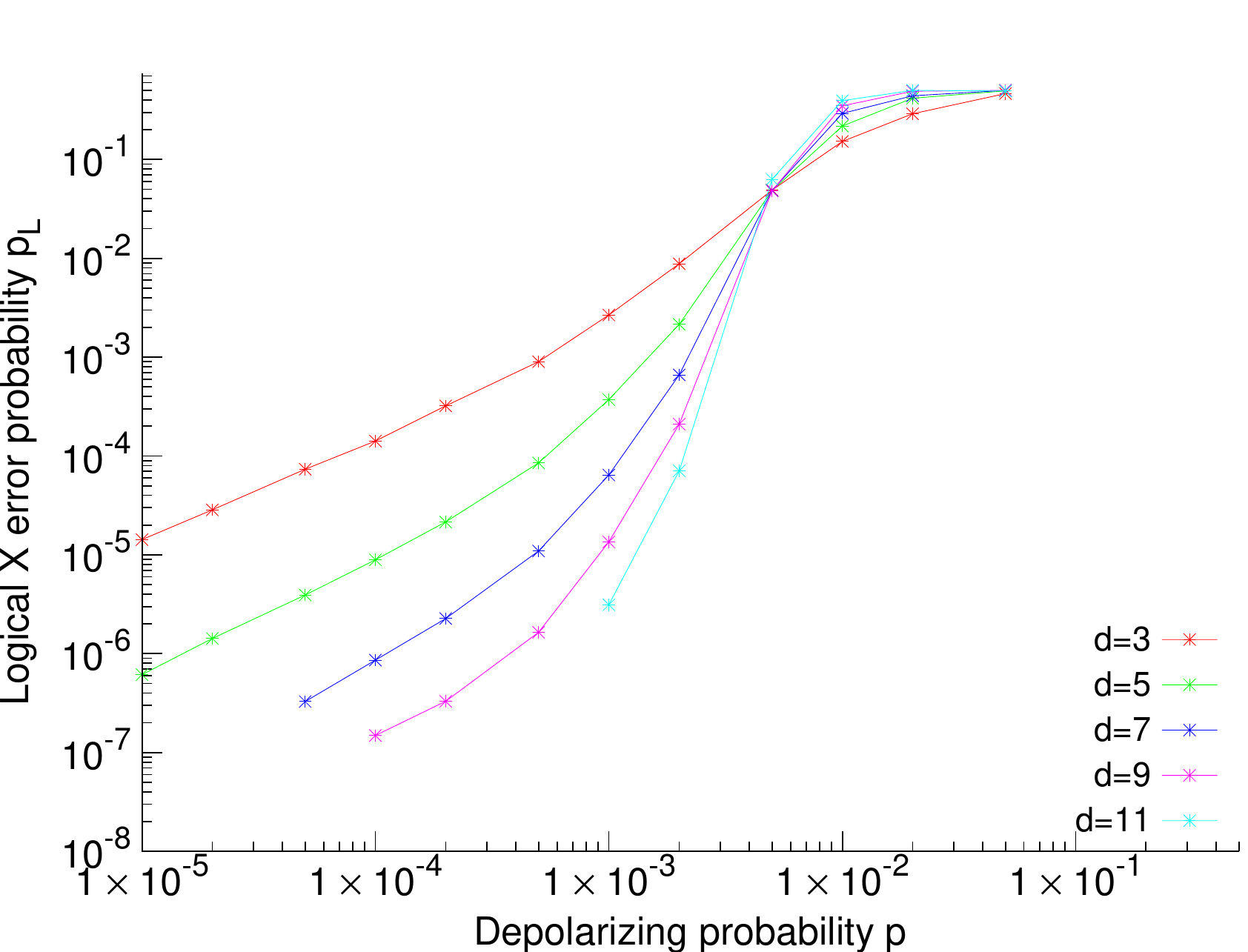}};
    \node [below right] at (9.5,6) {d)};
    \node [below right] at (11,6) {$n=5$};
\end{tikzpicture}
\end{center}
\caption{(Color online) Probability of surface code logical $X$ error per round of error detection for various code distances $d$ and physical error rates $p$ when increasingly large area errors are polynomially suppressed as a) $r^2/10$, b) $r^3/10$, c) $r^4/10$, d) $r^5/10$. Referring to the left of each graph, distance increases top to bottom. When suppression is quadratic, arbitrarily low logical error rates cannot be achieved at any finite value of $p$. For higher order polynomial suppression, arbitrarily low logical error rates can be achieved, however logical error is only suppressed polynomially with code distance, which may in some cases lead to unacceptable qubit overhead.}\label{poly}
\end{figure*}
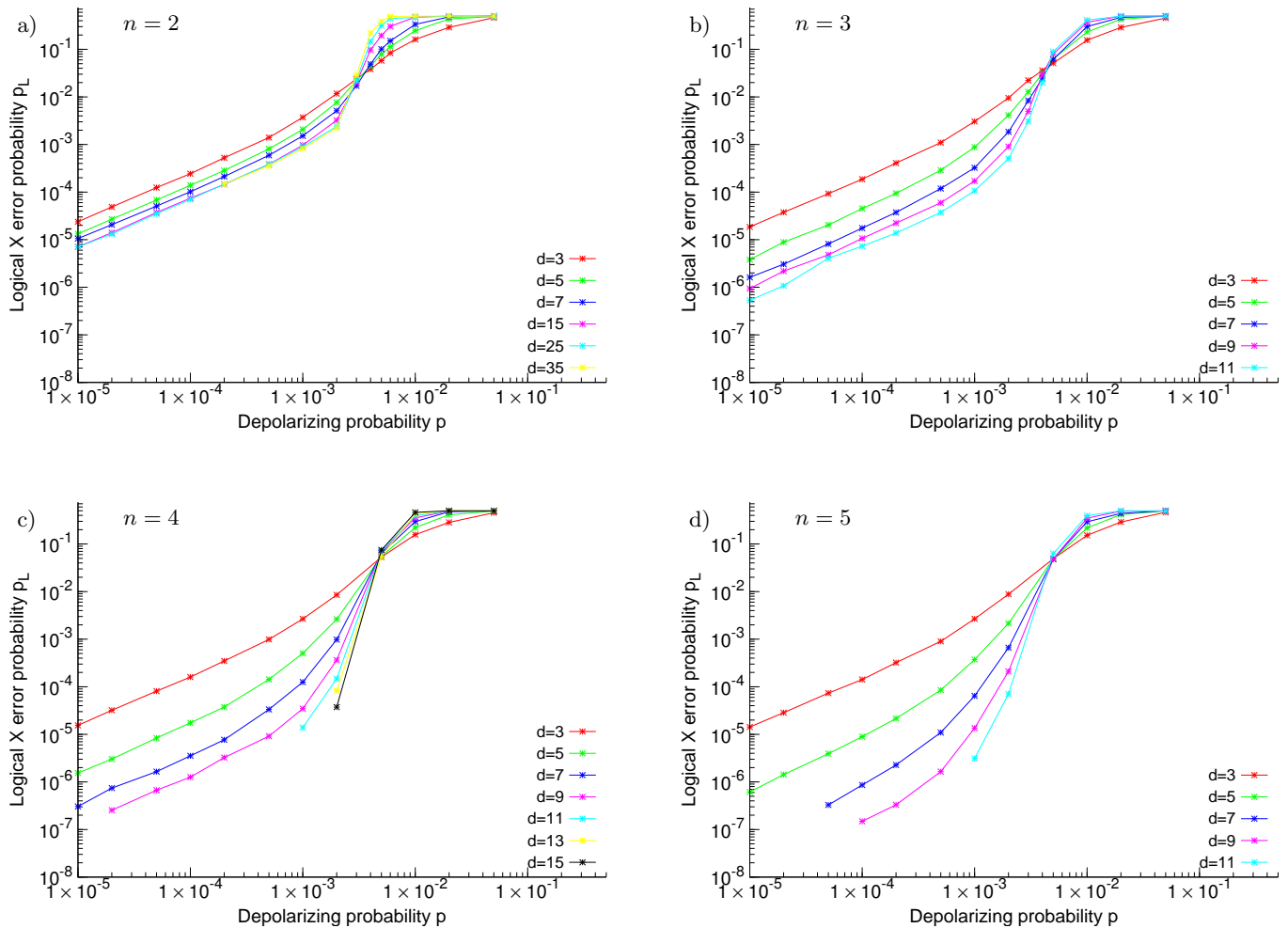

We now consider polynomial suppression of large-area errors (Fig.~\ref{poly}). If large-area errors are only quadratically suppressed, adding an additional ring of qubits at distance $r$ from any given qubit adds an $O(1/r)$ amount of error to that qubit, hence larger lattices of qubits will always be more error-prone and no threshold error rate will exist. For any rate of suppression greater than quadratic, arbitrarily reliable quantum computation can be achieved in principle, as the total error seen by any given qubit in an infinite lattice of qubits is bounded by a multiple of $p$.

At a moderately high error rate such as $p=10^{-3}$ and modest code distances, the dominant logical error contribution is from multiple temporally local errors. Such logical errors are exponentially suppressed with increasing code distance. To be explicit, for $n=4$, the polynomial of best fit through the data at $p=10^{-3}$ is order 8 in $d$, and for $n=5$ the best fit polynomial is order 16, clearly demonstrating that, in the high $p$ low $d$ regime, logical errors from single large-area physical errors are not dominant. At very large code distances, the quadratic growth of the number of gates per round of error detection and the exponential suppression of logical errors from multiple temporally local gate errors is expected to lead to weak $O(1/d^{n-2})$ suppression of logical error due to single very large area errors, however this regime is outside what we can currently reach with simulations.

Based only on currently accessible parameter ranges, at $p=10^{-3}$ the polynomial $n=4$ and $n=5$ overhead to achieve a given logical error rate is similar to the exponential $n=10$ overhead. If the computation being protected by the surface code is not too large, it therefore may well be the case that the desired logical error rate can be reached without excessive overhead with only polynomial suppression of large-area errors at the physical level. Formally, however, it should be noted that the resources required to achieve computation with logical error $\epsilon$ would grow polynomially with $1/\epsilon$ for sufficiently small $\epsilon$, which is not efficient in the computer science sense.

\section{Surface code performance with non-local two-qubit errors}
\label{2q}

Only two-body interactions are observed in nature between fundamental particles, meaning the large-area multi-qubit errors considered in the previous Section could only arise from uncontrolled \emph{engineered} multi-qubit interactions within a quantum computer or other exotic effects such as the radiation heating model described in Section~\ref{corr}. Unwanted two-body interactions, such as uncompensated Coulomb or magnetic dipole interaction, give rise to qualitatively and quantitatively different behavior. In this Section, we shall focus on long-range effects, and will therefore not consider interactions that decay exponentially quickly. As we shall see, even weakly polynomially decaying long-range interactions are quite tolerable, further justifying not considering exponentially decaying two-body interactions.

\begin{figure}
\begin{center}
\begin{tikzpicture}
    \node [anchor=south west, inner sep=0] at (0,7) {\includegraphics[width=85mm, viewport=60 60 545 430, clip=true]{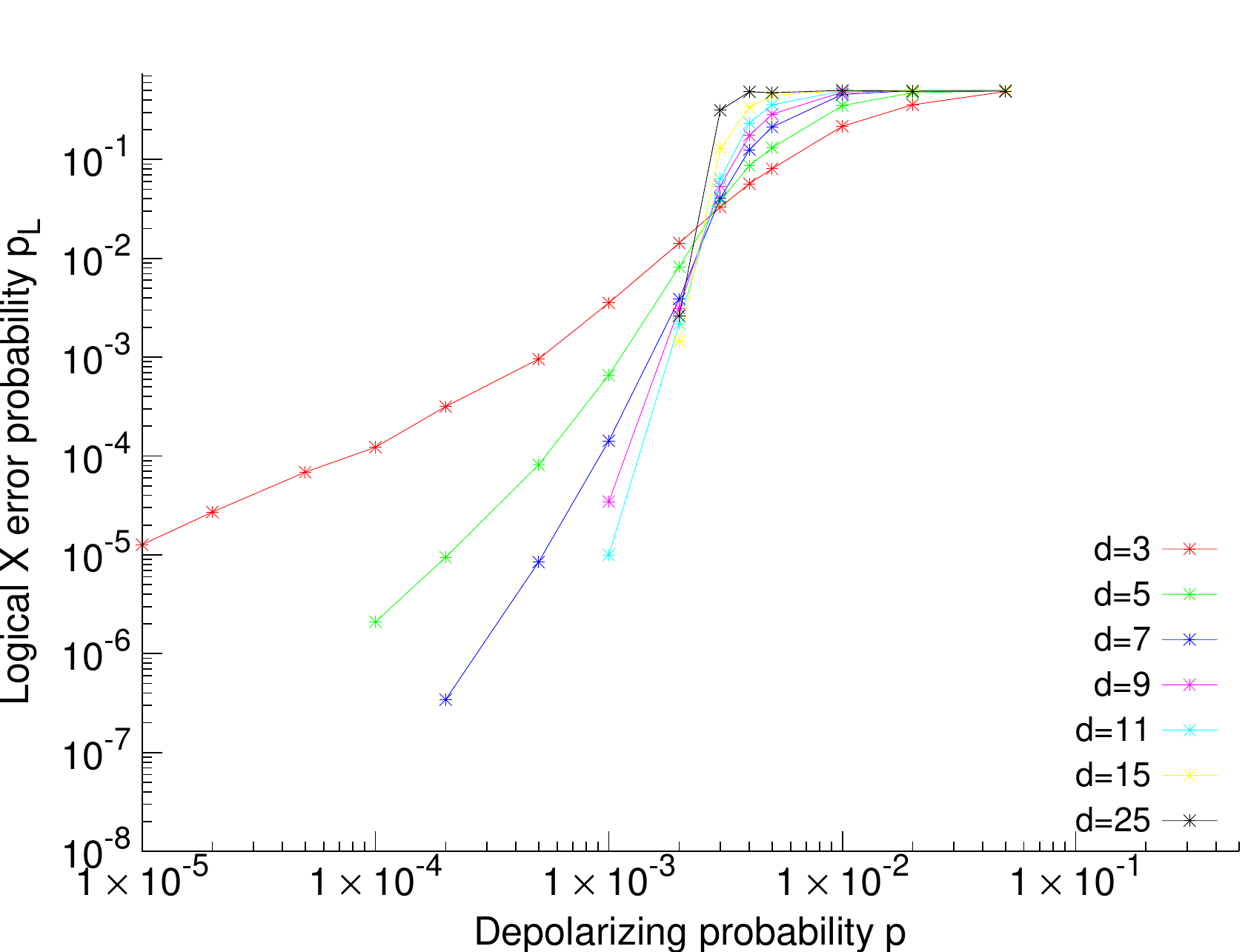}};
    \node [below right] at (0,13) {a)};
    \node [below right] at (1.5,13) {$p/r^2$};

    \node [anchor=south west, inner sep=0] at (0,0) {\includegraphics[width=85mm, viewport=60 60 545 430, clip=true]{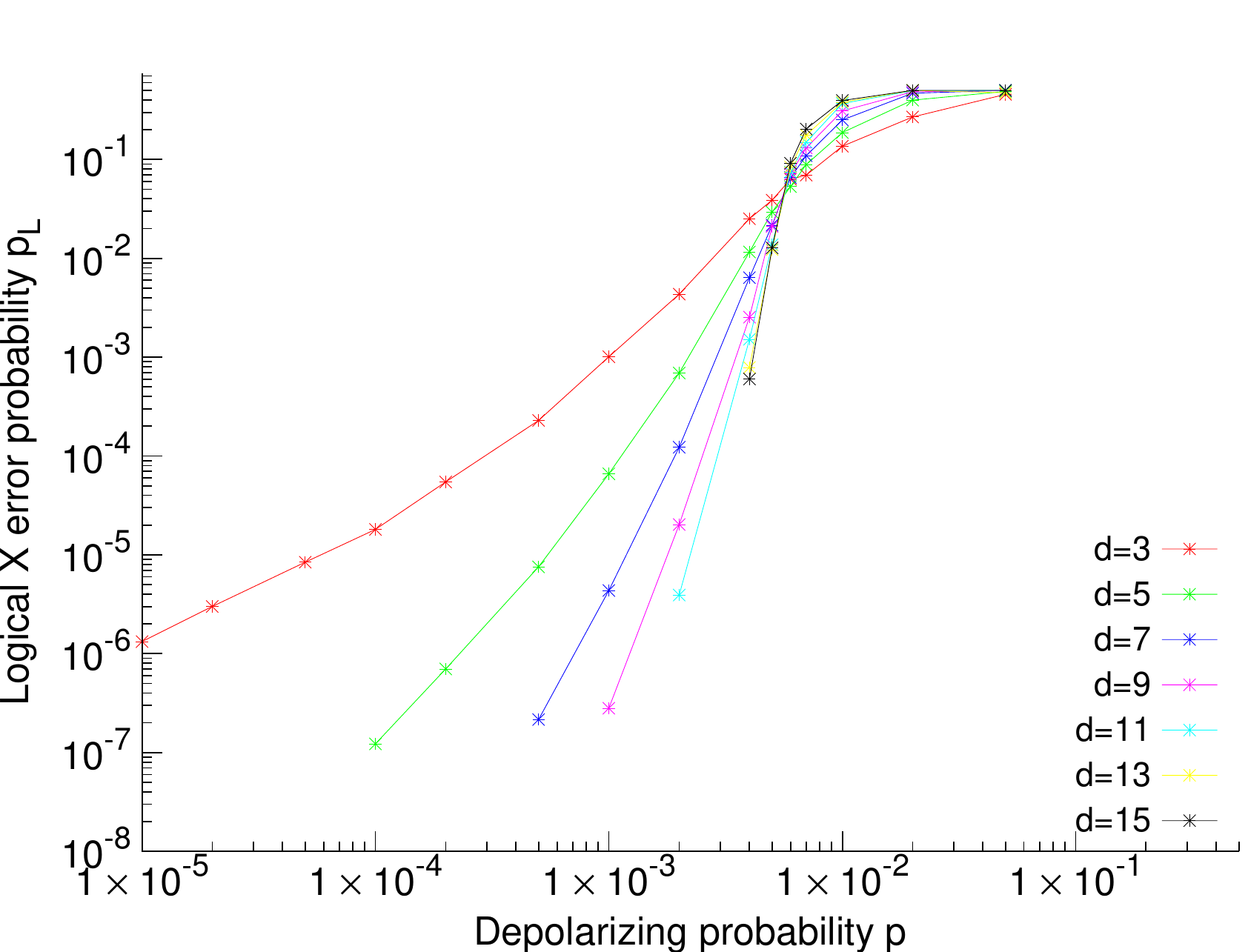}};
    \node [below right] at (0,6) {b)};
    \node [below right] at (1.5,6) {$0.1p/r^2$};
\end{tikzpicture}
\end{center}
\caption{(Color online) Probability of surface code logical $X$ error per round of error detection for various code distances $d$ and physical error rates $p$ when all pairs of qubits suffered two-qubit noise once per round of error detection with probability a) $p/r^2$, and b) $0.1p/r^2$. Referring to the left of each graph, distance increases top to bottom.}\label{poly2q}
\end{figure}

Any interaction between qubits is a potential source of unwanted evolution and hence error. When simulating the surface code using an array of qubits with polynomially decaying two-body interactions, if the characteristic gate error rate is $p$, at the beginning of each round of error detection each pair of qubits shall be modeled as suffering two-qubit depolarizing noise with probability $Ap/r^n$. We shall focus on the most severe $n=2$ case, and two values $A=1$ and $A=0.1$. The performance of the surface code with these two different levels of additional noise is shown in Fig.~\ref{poly2q}.

It should be stressed that, as with quadratically suppressed large-area errors, any qubit in an infinite 2-D lattice of qubits will suffer unbounded error and the surface code will fail. However, it can be seen that for the finite-size qubit arrays considered in simulations, robust suppression of logical error can be achieved even for the most severe $A=1$ case. The effect of a lack of a threshold error rate can be observed at $p=2\times 10^{-3}$ where the $d=25$ logical error rate is higher than that for $d=11$. Nevertheless, at error rates $p<10^{-3}$, the observed logical error rate suppression trend with increasing code distance suggests that extremely low logical error rates can be achieved before using larger code distances starts to hurt. Note that for $A=1$ and $p\leq 5\times 10^{-4}$ the observed logical error rates are less than or equal to those observed for $n=10$ exponential large-area errors, meaning the overhead will be less than the factor of 2.5 calculated in the previous Section, for moderate values of $d$.

By making the computer quasi 2-D, namely a finite width 1-D strip, the physical error seen at any given qubit would only grow logarithmically with increasing strip length, very likely permitting a usefully large number of logical qubits with usefully low logical error rates to be achieved. Other techniques such as building an array with carefully arranged walls capable of shielding the problematic interaction, or coupling widely separated finite arrays with other types of quantum communication are also possible. In short, even severe long-range two-qubit quantum errors that are only suppressed quadratically with increasing qubit separation can be handled with practical overhead.

\begin{figure}
\begin{center}
\begin{tikzpicture}
    \node [anchor=south west, inner sep=0] at (0,7) {\includegraphics[width=85mm, viewport=60 60 545 430, clip=true]{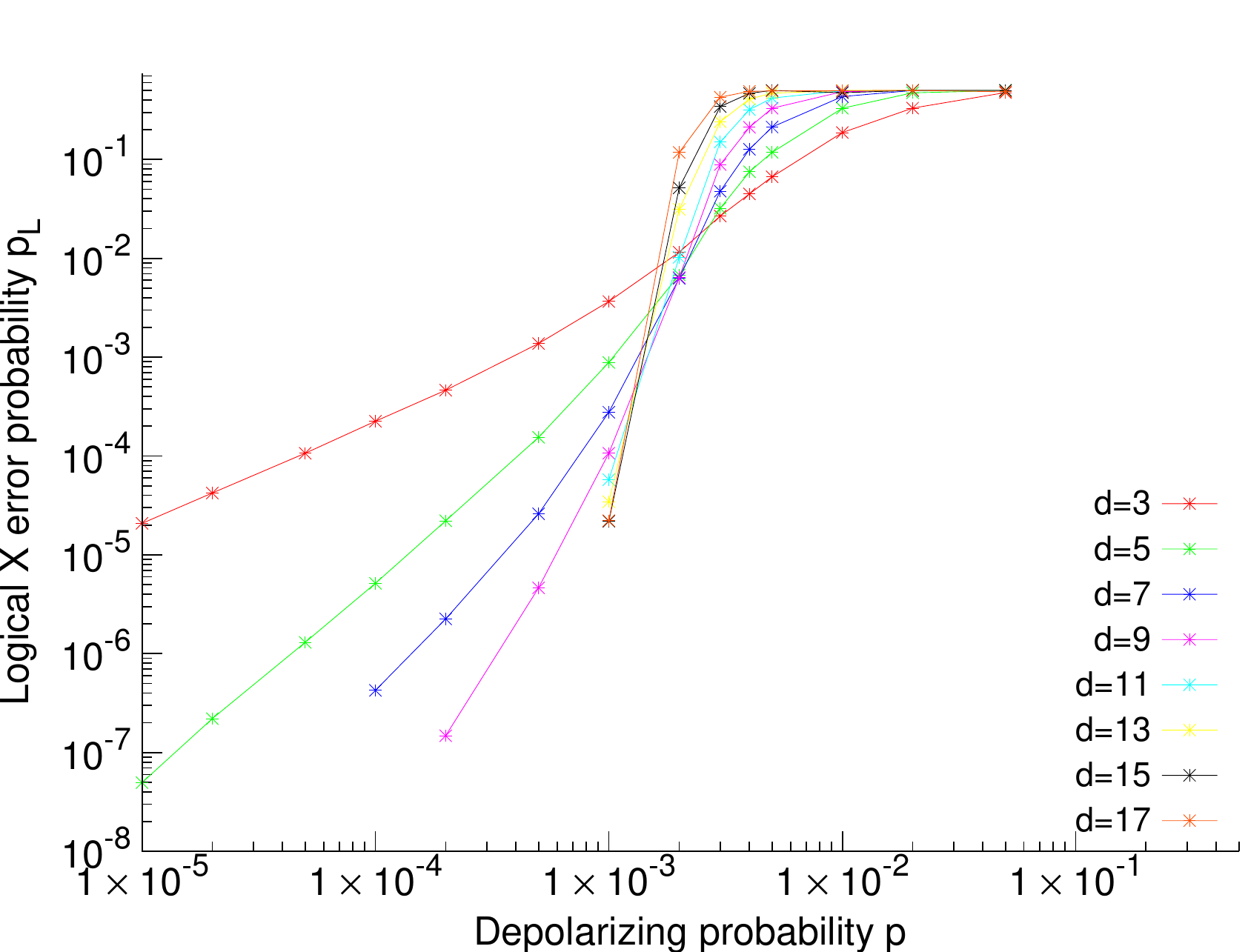}};
    \node [below right] at (0,13) {a)};
    \node [below right] at (1.5,13) {$p$};

    \node [anchor=south west, inner sep=0] at (0,0) {\includegraphics[width=85mm, viewport=60 60 545 430, clip=true]{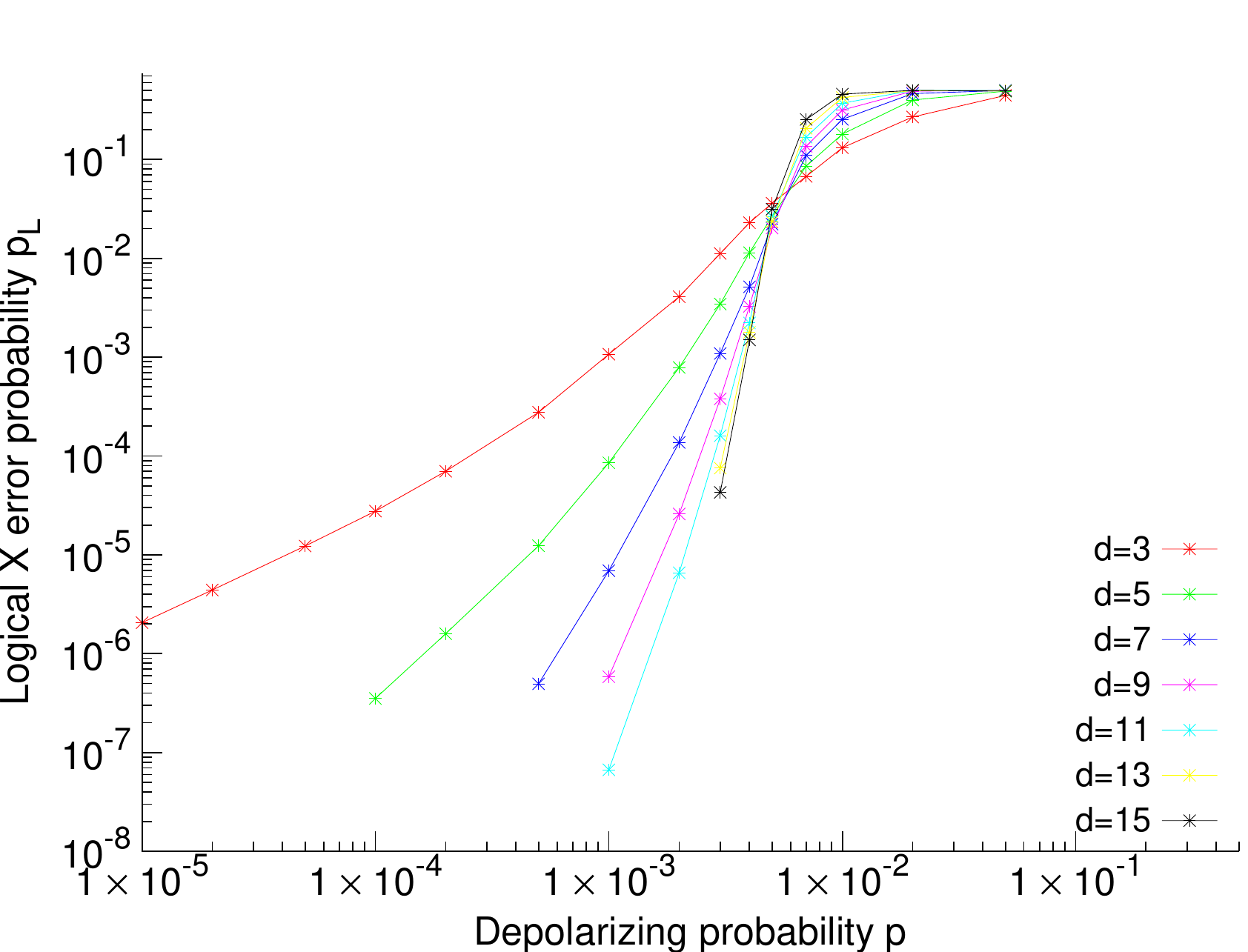}};
    \node [below right] at (0,6) {b)};
    \node [below right] at (1.5,6) {$0.1p$};
\end{tikzpicture}
\end{center}
\caption{(Color online) Probability of surface code logical $X$ error per round of error detection for various code distances $d$ and physical error rates $p$ when all pairs of qubits in each column of the surface code suffer two-qubit noise once per round of error detection with probability a) $p$, and b) $0.1p$. Referring to the left of each graph, distance increases top to bottom.}\label{const2q}
\end{figure}

The final class of error we shall consider are those arising from large-scale coupling elements that interact with many qubits, specifically entire columns of the surface code in the situation we shall model. Our basic motivating system is a chain of spins in a global magnetic field and shared inductive loop. Any pair of antiparallel spins can spontaneously flip, so we shall model this as a probability $Ap$ of error for every qubit pair in each column. Note that there is no suppression of this error with increasing qubit separation. Since any given qubit in a column has an increasing number of potential partners to flip with as the size of the surface code grows, there will again be no formal threshold error rate. We again focus on $A=1$ and $A=0.1$. Data is shown in Fig.~\ref{const2q}.

For $A=1$ (Fig.~\ref{const2q}a), it can be seen that at $p=10^{-3}$ the lowest possible logical error rate is achieved with a distance 15 code. Fig.~\ref{poly2q} and Fig.~\ref{const2q} are qualitatively very similar as in both an array of qubits with quadratically suppressed interactions between all pairs of qubits and an array of qubits with columns coupled by single devices introducing errors with no suppression with increasing distance, the total error seen by any given qubit grows linearly with code distance. When $A=0.1$ (Fig.~\ref{const2q}b), at $p=10^{-3}$ it can be seen that very low logical error rates  can be achieved with modest code distances. Again, despite the lack of a threshold error rate, it can be seen that this class of errors is tolerable with low overhead in practice.

\section{Conclusion}
\label{conc}

We have shown that moderate exponential suppression of large-area errors is sufficient to observe strong exponential suppression of logical error with increasing code distance $d$. A factor of 10 suppression of each successively larger area class of physical errors leads to just a factor of 2.5 additional qubits to achieve the same logical error observed without large-area errors when gates have characteristic error $p=10^{-3}$. Overhead is negligible ($<$10\%) for moderately large algorithm sizes and a factor of suppression of $10^3$. Since 5+ body errors are expected to be exceedingly rare in most physical setups, it is reasonable to expect that this higher level of error suppression is experimentally achievable and that large-area errors can therefore mostly be ignored when analyzing the surface code.

A second class of errors, namely long-range two-qubit errors, has been shown to be remarkably tolerable, with even lower overhead than the exponentially suppressed large-area errors for $p\lesssim 10^{-3}$.  This is surprising as such noise, from a formal point of view, results in no threshold error rate, meaning arbitrarily reliable quantum computation cannot be achieved at any finite error rate. Nevertheless, sufficiently low logical error rates for practical purposes can be achieved with modest code distances.

In all cases where a threshold error rate exists, it remains well above $10^{-3}$ and in most cases does not stray far from the baseline threshold error rate of approximately 0.5\%. This is in line with expectations as the correlated errors introduced in the simulations are typically at least an order of magnitude less likely than the baseline gate errors, meaning they have low impact around the threshold error rate. The only exception to this is $n=2$ exponentially suppressed large-area errors, where weight 5 errors, for example, are only half as likely as single-qubit errors, resulting in a degradation of the threshold error rate to just above $10^{-3}$.

Collectively, these results imply that large-area and long-range errors pose no fundamental barriers to practical large-scale quantum computation, as both classes of error, from a practical point of view, can be well handled by the surface code. Experimentally, the implication is that, in a large device, one should focus on the gate error rate observed when the maximum possible number of qubits in the array are being actively manipulated in parallel. The parallel error rate is the figure of merit required to determine whether a physical device can be used to achieve low logical error rates.

\section{Acknowledgements}
\label{ack}

We thank Daniel Gottesman for suggesting this project, and Rami Barends, Julian Kelly, Daniel Sank, Evan Jeffrey, Ted White, and John Preskill for helpful discussions. This research was funded by the US Office of the Director of National Intelligence (ODNI), Intelligence Advanced Research Projects Activity (IARPA), through the US Army Research Office grant No. W911NF-10-1-0334. Supported in part by the Australian Research Council Centre of Excellence for Quantum Computation and Communication Technology (CE110001027) and the U.S. Army Research Office (W911NF-13-1-0024). All statements of fact, opinion or conclusions contained herein are those of the authors and should not be construed as representing the official views or policies of IARPA, the ODNI, or the US Government.

\bibliography{../References}

\begin{thebibliography}{21}%
\makeatletter
\providecommand \@ifxundefined [1]{%
 \@ifx{#1\undefined}
}%
\providecommand \@ifnum [1]{%
 \ifnum #1\expandafter \@firstoftwo
 \else \expandafter \@secondoftwo
 \fi
}%
\providecommand \@ifx [1]{%
 \ifx #1\expandafter \@firstoftwo
 \else \expandafter \@secondoftwo
 \fi
}%
\providecommand \natexlab [1]{#1}%
\providecommand \enquote  [1]{``#1''}%
\providecommand \bibnamefont  [1]{#1}%
\providecommand \bibfnamefont [1]{#1}%
\providecommand \citenamefont [1]{#1}%
\providecommand \href@noop [0]{\@secondoftwo}%
\providecommand \href [0]{\begingroup \@sanitize@url \@href}%
\providecommand \@href[1]{\@@startlink{#1}\@@href}%
\providecommand \@@href[1]{\endgroup#1\@@endlink}%
\providecommand \@sanitize@url [0]{\catcode `\\12\catcode `\$12\catcode
  `\&12\catcode `\#12\catcode `\^12\catcode `\_12\catcode `\%12\relax}%
\providecommand \@@startlink[1]{}%
\providecommand \@@endlink[0]{}%
\providecommand \url  [0]{\begingroup\@sanitize@url \@url }%
\providecommand \@url [1]{\endgroup\@href {#1}{\urlprefix }}%
\providecommand \urlprefix  [0]{URL }%
\providecommand \Eprint [0]{\href }%
\providecommand \doibase [0]{http://dx.doi.org/}%
\providecommand \selectlanguage [0]{\@gobble}%
\providecommand \bibinfo  [0]{\@secondoftwo}%
\providecommand \bibfield  [0]{\@secondoftwo}%
\providecommand \translation [1]{[#1]}%
\providecommand \BibitemOpen [0]{}%
\providecommand \bibitemStop [0]{}%
\providecommand \bibitemNoStop [0]{.\EOS\space}%
\providecommand \EOS [0]{\spacefactor3000\relax}%
\providecommand \BibitemShut  [1]{\csname bibitem#1\endcsname}%
\let\auto@bib@innerbib\@empty
\bibitem [{\citenamefont {Nayak}\ \emph {et~al.}(2008)\citenamefont {Nayak},
  \citenamefont {Simon}, \citenamefont {Stern}, \citenamefont {Freedman},\ and\
  \citenamefont {Sarma}}]{Naya08}%
  \BibitemOpen
  \bibfield  {author} {\bibinfo {author} {\bibfnamefont {C.}~\bibnamefont
  {Nayak}}, \bibinfo {author} {\bibfnamefont {S.~H.}\ \bibnamefont {Simon}},
  \bibinfo {author} {\bibfnamefont {A.}~\bibnamefont {Stern}}, \bibinfo
  {author} {\bibfnamefont {M.}~\bibnamefont {Freedman}}, \ and\ \bibinfo
  {author} {\bibfnamefont {S.~D.}\ \bibnamefont {Sarma}},\ }\href@noop {}
  {\bibfield  {journal} {\bibinfo  {journal} {Rev. Mod. Phys.}\ }\textbf
  {\bibinfo {volume} {80}},\ \bibinfo {pages} {1083} (\bibinfo {year}
  {2008})},\ \bibinfo {note} {arXiv:0707.1889}\BibitemShut {NoStop}%
\bibitem [{\citenamefont {Bonesteel}\ and\ \citenamefont
  {DiVincenzo}(2012)}]{Bone12}%
  \BibitemOpen
  \bibfield  {author} {\bibinfo {author} {\bibfnamefont {N.~E.}\ \bibnamefont
  {Bonesteel}}\ and\ \bibinfo {author} {\bibfnamefont {D.~P.}\ \bibnamefont
  {DiVincenzo}},\ }\href@noop {} {\bibfield  {journal} {\bibinfo  {journal}
  {Phys. Rev. B}\ }\textbf {\bibinfo {volume} {86}},\ \bibinfo {pages} {165113}
  (\bibinfo {year} {2012})},\ \bibinfo {note} {arXiv:1206.6048}\BibitemShut
  {NoStop}%
\bibitem [{\citenamefont {Gottesman}(2013)}]{Gott13}%
  \BibitemOpen
  \bibfield  {author} {\bibinfo {author} {\bibfnamefont {D.}~\bibnamefont
  {Gottesman}},\ }\href@noop {} {\bibfield  {journal} {\bibinfo  {journal}
  {arXiv:1310.2984}\ } (\bibinfo {year} {2013})}\BibitemShut {NoStop}%
\bibitem [{\citenamefont {Bombin}(2013)}]{Bomb13}%
  \BibitemOpen
  \bibfield  {author} {\bibinfo {author} {\bibfnamefont {H.}~\bibnamefont
  {Bombin}},\ }\href@noop {} {\bibfield  {journal} {\bibinfo  {journal}
  {arXiv:1311.0879}\ } (\bibinfo {year} {2013})}\BibitemShut {NoStop}%
\bibitem [{\citenamefont {Bravyi}\ and\ \citenamefont
  {Hastings}(2013)}]{Brav13b}%
  \BibitemOpen
  \bibfield  {author} {\bibinfo {author} {\bibfnamefont {S.}~\bibnamefont
  {Bravyi}}\ and\ \bibinfo {author} {\bibfnamefont {M.~B.}\ \bibnamefont
  {Hastings}},\ }\href@noop {} {\bibfield  {journal} {\bibinfo  {journal}
  {arXiv:1311.0885}\ } (\bibinfo {year} {2013})}\BibitemShut {NoStop}%
\bibitem [{\citenamefont {Bravyi}\ and\ \citenamefont {Kitaev}(1998)}]{Brav98}%
  \BibitemOpen
  \bibfield  {author} {\bibinfo {author} {\bibfnamefont {S.~B.}\ \bibnamefont
  {Bravyi}}\ and\ \bibinfo {author} {\bibfnamefont {A.~Y.}\ \bibnamefont
  {Kitaev}},\ }\href@noop {} {\bibfield  {journal} {\bibinfo  {journal}
  {quant-ph/9811052}\ } (\bibinfo {year} {1998})}\BibitemShut {NoStop}%
\bibitem [{\citenamefont {Dennis}(2001)}]{Denn99}%
  \BibitemOpen
  \bibfield  {author} {\bibinfo {author} {\bibfnamefont {E.}~\bibnamefont
  {Dennis}},\ }\href@noop {} {\bibfield  {journal} {\bibinfo  {journal} {Phys.
  Rev. A}\ }\textbf {\bibinfo {volume} {63}},\ \bibinfo {pages} {052314}
  (\bibinfo {year} {2001})},\ \bibinfo {note} {quant-ph/9905027}\BibitemShut
  {NoStop}%
\bibitem [{\citenamefont {Raussendorf}\ and\ \citenamefont
  {Harrington}(2007)}]{Raus07}%
  \BibitemOpen
  \bibfield  {author} {\bibinfo {author} {\bibfnamefont {R.}~\bibnamefont
  {Raussendorf}}\ and\ \bibinfo {author} {\bibfnamefont {J.}~\bibnamefont
  {Harrington}},\ }\href@noop {} {\bibfield  {journal} {\bibinfo  {journal}
  {Phys. Rev. Lett.}\ }\textbf {\bibinfo {volume} {98}},\ \bibinfo {pages}
  {190504} (\bibinfo {year} {2007})},\ \bibinfo {note}
  {quant-ph/0610082}\BibitemShut {NoStop}%
\bibitem [{\citenamefont {Raussendorf}\ \emph {et~al.}(2007)\citenamefont
  {Raussendorf}, \citenamefont {Harrington},\ and\ \citenamefont
  {Goyal}}]{Raus07d}%
  \BibitemOpen
  \bibfield  {author} {\bibinfo {author} {\bibfnamefont {R.}~\bibnamefont
  {Raussendorf}}, \bibinfo {author} {\bibfnamefont {J.}~\bibnamefont
  {Harrington}}, \ and\ \bibinfo {author} {\bibfnamefont {K.}~\bibnamefont
  {Goyal}},\ }\href@noop {} {\bibfield  {journal} {\bibinfo  {journal} {New J.
  Phys.}\ }\textbf {\bibinfo {volume} {9}},\ \bibinfo {pages} {199} (\bibinfo
  {year} {2007})},\ \bibinfo {note} {quant-ph/0703143}\BibitemShut {NoStop}%
\bibitem [{\citenamefont {Fowler}\ \emph {et~al.}(2012)\citenamefont {Fowler},
  \citenamefont {Mariantoni}, \citenamefont {Martinis},\ and\ \citenamefont
  {Cleland}}]{Fowl12f}%
  \BibitemOpen
  \bibfield  {author} {\bibinfo {author} {\bibfnamefont {A.~G.}\ \bibnamefont
  {Fowler}}, \bibinfo {author} {\bibfnamefont {M.}~\bibnamefont {Mariantoni}},
  \bibinfo {author} {\bibfnamefont {J.~M.}\ \bibnamefont {Martinis}}, \ and\
  \bibinfo {author} {\bibfnamefont {A.~N.}\ \bibnamefont {Cleland}},\
  }\href@noop {} {\bibfield  {journal} {\bibinfo  {journal} {Phys. Rev. A}\
  }\textbf {\bibinfo {volume} {86}},\ \bibinfo {pages} {032324} (\bibinfo
  {year} {2012})},\ \bibinfo {note} {arXiv:1208.0928}\BibitemShut {NoStop}%
\bibitem [{\citenamefont {Barends}\ \emph {et~al.}()\citenamefont {Barends},
  \citenamefont {Kelly}, \citenamefont {Megrant}, \citenamefont {Veitia},
  \citenamefont {Sank}, \citenamefont {Jeffrey}, \citenamefont {White},
  \citenamefont {Mutus}, \citenamefont {Fowler}, \citenamefont {Campbell},
  \citenamefont {Chen}, \citenamefont {Chen}, \citenamefont {Chiaro},
  \citenamefont {Dunsworth}, \citenamefont {Neill}, \citenamefont {O'Malley},
  \citenamefont {Roushan}, \citenamefont {Vainsencher}, \citenamefont {Wenner},
  \citenamefont {Korotkov}, \citenamefont {Cleland}, ,\ and\ \citenamefont
  {Martinis}}]{Bare13}%
  \BibitemOpen
  \bibfield  {author} {\bibinfo {author} {\bibfnamefont {R.}~\bibnamefont
  {Barends}}, \bibinfo {author} {\bibfnamefont {J.}~\bibnamefont {Kelly}},
  \bibinfo {author} {\bibfnamefont {A.}~\bibnamefont {Megrant}}, \bibinfo
  {author} {\bibfnamefont {A.}~\bibnamefont {Veitia}}, \bibinfo {author}
  {\bibfnamefont {D.}~\bibnamefont {Sank}}, \bibinfo {author} {\bibfnamefont
  {E.}~\bibnamefont {Jeffrey}}, \bibinfo {author} {\bibfnamefont {T.~C.}\
  \bibnamefont {White}}, \bibinfo {author} {\bibfnamefont {J.}~\bibnamefont
  {Mutus}}, \bibinfo {author} {\bibfnamefont {A.~G.}\ \bibnamefont {Fowler}},
  \bibinfo {author} {\bibfnamefont {B.}~\bibnamefont {Campbell}}, \bibinfo
  {author} {\bibfnamefont {Y.}~\bibnamefont {Chen}}, \bibinfo {author}
  {\bibfnamefont {Z.}~\bibnamefont {Chen}}, \bibinfo {author} {\bibfnamefont
  {B.}~\bibnamefont {Chiaro}}, \bibinfo {author} {\bibfnamefont
  {A.}~\bibnamefont {Dunsworth}}, \bibinfo {author} {\bibfnamefont
  {C.}~\bibnamefont {Neill}}, \bibinfo {author} {\bibfnamefont
  {P.}~\bibnamefont {O'Malley}}, \bibinfo {author} {\bibfnamefont
  {P.}~\bibnamefont {Roushan}}, \bibinfo {author} {\bibfnamefont
  {A.}~\bibnamefont {Vainsencher}}, \bibinfo {author} {\bibfnamefont
  {J.}~\bibnamefont {Wenner}}, \bibinfo {author} {\bibfnamefont {A.~N.}\
  \bibnamefont {Korotkov}}, \bibinfo {author} {\bibfnamefont {A.~N.}\
  \bibnamefont {Cleland}}, , \ and\ \bibinfo {author} {\bibfnamefont {J.~M.}\
  \bibnamefont {Martinis}},\ }\href@noop {} {\ }\bibinfo {note}
  {Submitted}\BibitemShut {NoStop}%
\bibitem [{\citenamefont {Aharonov}\ \emph {et~al.}(2006)\citenamefont
  {Aharonov}, \citenamefont {Kitaev},\ and\ \citenamefont {Preskill}}]{Ahar06}%
  \BibitemOpen
  \bibfield  {author} {\bibinfo {author} {\bibfnamefont {D.}~\bibnamefont
  {Aharonov}}, \bibinfo {author} {\bibfnamefont {A.}~\bibnamefont {Kitaev}}, \
  and\ \bibinfo {author} {\bibfnamefont {J.}~\bibnamefont {Preskill}},\
  }\href@noop {} {\bibfield  {journal} {\bibinfo  {journal} {Phys. Rev. Lett.}\
  }\textbf {\bibinfo {volume} {96}},\ \bibinfo {pages} {050504} (\bibinfo
  {year} {2006})},\ \bibinfo {note} {quant-ph/0510231}\BibitemShut {NoStop}%
\bibitem [{\citenamefont {Ng}\ and\ \citenamefont {Preskill}(2009)}]{Ng09}%
  \BibitemOpen
  \bibfield  {author} {\bibinfo {author} {\bibfnamefont {H.~K.}\ \bibnamefont
  {Ng}}\ and\ \bibinfo {author} {\bibfnamefont {J.}~\bibnamefont {Preskill}},\
  }\href@noop {} {\bibfield  {journal} {\bibinfo  {journal} {Phys. Rev. A}\
  }\textbf {\bibinfo {volume} {79}},\ \bibinfo {pages} {032318} (\bibinfo
  {year} {2009})},\ \bibinfo {note} {arXiv:0810.4953}\BibitemShut {NoStop}%
\bibitem [{\citenamefont {Preskill}(2013)}]{Pres13}%
  \BibitemOpen
  \bibfield  {author} {\bibinfo {author} {\bibfnamefont {J.}~\bibnamefont
  {Preskill}},\ }\href@noop {} {\bibfield  {journal} {\bibinfo  {journal}
  {Quant. Inf. Comput.}\ }\textbf {\bibinfo {volume} {13}},\ \bibinfo {pages}
  {181} (\bibinfo {year} {2013})},\ \bibinfo {note}
  {arXiv:1207.6131}\BibitemShut {NoStop}%
\bibitem [{\citenamefont {Barends}\ \emph {et~al.}(2013)\citenamefont
  {Barends}, \citenamefont {Kelly}, \citenamefont {Megrant}, \citenamefont
  {Sank}, \citenamefont {Jeffrey}, \citenamefont {Chen}, \citenamefont {Yin},
  \citenamefont {Chiaro}, \citenamefont {Mutus}, \citenamefont {Neill},
  \citenamefont {O'Malley}, \citenamefont {Roushan}, \citenamefont {Wenner},
  \citenamefont {White}, \citenamefont {Cleland},\ and\ \citenamefont
  {Martinis}}]{Bare13a}%
  \BibitemOpen
  \bibfield  {author} {\bibinfo {author} {\bibfnamefont {R.}~\bibnamefont
  {Barends}}, \bibinfo {author} {\bibfnamefont {J.}~\bibnamefont {Kelly}},
  \bibinfo {author} {\bibfnamefont {A.}~\bibnamefont {Megrant}}, \bibinfo
  {author} {\bibfnamefont {D.}~\bibnamefont {Sank}}, \bibinfo {author}
  {\bibfnamefont {E.}~\bibnamefont {Jeffrey}}, \bibinfo {author} {\bibfnamefont
  {Y.}~\bibnamefont {Chen}}, \bibinfo {author} {\bibfnamefont {Y.}~\bibnamefont
  {Yin}}, \bibinfo {author} {\bibfnamefont {B.}~\bibnamefont {Chiaro}},
  \bibinfo {author} {\bibfnamefont {J.}~\bibnamefont {Mutus}}, \bibinfo
  {author} {\bibfnamefont {C.}~\bibnamefont {Neill}}, \bibinfo {author}
  {\bibfnamefont {P.}~\bibnamefont {O'Malley}}, \bibinfo {author}
  {\bibfnamefont {P.}~\bibnamefont {Roushan}}, \bibinfo {author} {\bibfnamefont
  {J.}~\bibnamefont {Wenner}}, \bibinfo {author} {\bibfnamefont {T.~C.}\
  \bibnamefont {White}}, \bibinfo {author} {\bibfnamefont {A.~N.}\ \bibnamefont
  {Cleland}}, \ and\ \bibinfo {author} {\bibfnamefont {J.~M.}\ \bibnamefont
  {Martinis}},\ }\href@noop {} {\bibfield  {journal} {\bibinfo  {journal}
  {Phys. Rev. Lett.}\ }\textbf {\bibinfo {volume} {111}},\ \bibinfo {pages}
  {080502} (\bibinfo {year} {2013})},\ \bibinfo {note}
  {arXiv:1304.2322}\BibitemShut {NoStop}%
\bibitem [{\citenamefont {Hollenberg}\ \emph {et~al.}(2006)\citenamefont
  {Hollenberg}, \citenamefont {Greentree}, \citenamefont {Fowler},\ and\
  \citenamefont {Wellard}}]{Holl06}%
  \BibitemOpen
  \bibfield  {author} {\bibinfo {author} {\bibfnamefont {L.~C.~L.}\
  \bibnamefont {Hollenberg}}, \bibinfo {author} {\bibfnamefont {A.~D.}\
  \bibnamefont {Greentree}}, \bibinfo {author} {\bibfnamefont {A.~G.}\
  \bibnamefont {Fowler}}, \ and\ \bibinfo {author} {\bibfnamefont {C.~J.}\
  \bibnamefont {Wellard}},\ }\href@noop {} {\bibfield  {journal} {\bibinfo
  {journal} {Phys. Rev. B}\ }\textbf {\bibinfo {volume} {74}},\ \bibinfo
  {pages} {045311} (\bibinfo {year} {2006})},\ \bibinfo {note}
  {quant-ph/0506198}\BibitemShut {NoStop}%
\bibitem [{\citenamefont {Loss}\ and\ \citenamefont
  {DiVincenzo}(1998)}]{Loss98}%
  \BibitemOpen
  \bibfield  {author} {\bibinfo {author} {\bibfnamefont {D.}~\bibnamefont
  {Loss}}\ and\ \bibinfo {author} {\bibfnamefont {D.~P.}\ \bibnamefont
  {DiVincenzo}},\ }\href@noop {} {\bibfield  {journal} {\bibinfo  {journal}
  {Phys. Rev. A}\ }\textbf {\bibinfo {volume} {57}},\ \bibinfo {pages} {120–}
  (\bibinfo {year} {1998})},\ \bibinfo {note} {cond-mat/9701055}\BibitemShut
  {NoStop}%
\bibitem [{\citenamefont {Kielpinski}\ \emph {et~al.}(2002)\citenamefont
  {Kielpinski}, \citenamefont {Monroe},\ and\ \citenamefont
  {Wineland}}]{Kiel02}%
  \BibitemOpen
  \bibfield  {author} {\bibinfo {author} {\bibfnamefont {D.}~\bibnamefont
  {Kielpinski}}, \bibinfo {author} {\bibfnamefont {C.}~\bibnamefont {Monroe}},
  \ and\ \bibinfo {author} {\bibfnamefont {D.~J.}\ \bibnamefont {Wineland}},\
  }\href@noop {} {\bibfield  {journal} {\bibinfo  {journal} {Nature}\ }\textbf
  {\bibinfo {volume} {417}},\ \bibinfo {pages} {709} (\bibinfo {year}
  {2002})}\BibitemShut {NoStop}%
\bibitem [{\citenamefont {Brennen}\ \emph {et~al.}(1999)\citenamefont
  {Brennen}, \citenamefont {Caves}, \citenamefont {Jessen},\ and\ \citenamefont
  {Deutsch}}]{Bren99}%
  \BibitemOpen
  \bibfield  {author} {\bibinfo {author} {\bibfnamefont {G.~K.}\ \bibnamefont
  {Brennen}}, \bibinfo {author} {\bibfnamefont {C.~M.}\ \bibnamefont {Caves}},
  \bibinfo {author} {\bibfnamefont {P.~S.}\ \bibnamefont {Jessen}}, \ and\
  \bibinfo {author} {\bibfnamefont {I.~H.}\ \bibnamefont {Deutsch}},\
  }\href@noop {} {\bibfield  {journal} {\bibinfo  {journal} {Phys. Rev. Lett.}\
  }\textbf {\bibinfo {volume} {82}},\ \bibinfo {pages} {1060} (\bibinfo {year}
  {1999})},\ \bibinfo {note} {quant-ph/9806021}\BibitemShut {NoStop}%
\bibitem [{\citenamefont {Misra}\ and\ \citenamefont
  {Sudarshan}(1977)}]{Misr77}%
  \BibitemOpen
  \bibfield  {author} {\bibinfo {author} {\bibfnamefont {B.}~\bibnamefont
  {Misra}}\ and\ \bibinfo {author} {\bibfnamefont {E.~C.~G.}\ \bibnamefont
  {Sudarshan}},\ }\href@noop {} {\bibfield  {journal} {\bibinfo  {journal} {J.
  Math. Phys.}\ }\textbf {\bibinfo {volume} {18}},\ \bibinfo {pages} {756}
  (\bibinfo {year} {1977})}\BibitemShut {NoStop}%
\bibitem [{\citenamefont {Fowler}(2013)}]{Fowl13g}%
  \BibitemOpen
  \bibfield  {author} {\bibinfo {author} {\bibfnamefont {A.~G.}\ \bibnamefont
  {Fowler}},\ }\href@noop {} {\bibfield  {journal} {\bibinfo  {journal}
  {arXiv:1310.0863}\ } (\bibinfo {year} {2013})}\BibitemShut {NoStop}%
\end{thebibliography}%

\end{document}